# Misfit stabilized embedded nanoparticles in metallic alloys

Yu. N. Gornostyrev and M. I. Katsnelson

Nanoscale inhomogeneities are typical for numerous metallic alloys and crucially important for their practical applications. At the same time, stabilization mechanisms of such a state are poorly understood. We present a general overview of the problem, together with a more detailed discussion of the prototype example, namely, Guinier-Preston zones in Al-based alloys. It is shown that coherent strain due to a misfit between inclusion and host crystal lattices plays a decisive role in the emergence of the inhomogeneous state. We suggest a model explaining formation of ultrathin plates (with the thickness of a few lattice constants) typical for Al-Cu alloys. Discreteness of the array of misfit dislocations and long-ranged elastic interactions between them are the key ingredients of the model. This opens a way to a general understanding of the nature of (meta)stable embedded nanoparticles in practically important systems.

## Introduction

One of the basic concepts in physics and chemistry of solids and materials science is the formation of microstructures which depend of chemical composition and thermal treatment conditions [1]. Of particular interest are microstructures with nanoscale elements [2] (tens to thousands interatomic distances) which properties differ essentially from those of single molecules/atoms as well as of bulk materials [3]. Their peculiarities are generally explained in terms of high surface-to-volume ratio and/or size effect [4] in fundamental properties of materials (such as, e.g., size quantization of electron energy spectrum in embedded nanoparticles [5]). We are just in the very beginning of the way, and specific physical mechanisms responsible for formation and stability of nano-scale microstructures are still debatable [6].

A typical example is a microstructure formed by the quenching of a high temperature state which presents either structural inhomogeneities or products of phase decomposition. In the former case, a tweed-like microstructure with short-range crystallographic order arises, characteristic of quenched pre-transition state (see for example [7,8,9]). In the latter case, the kinetics of the first-order phase transition [10] is crucially important and the nanoscale microstructure is formed due to freezing/stabilization of an incomplete stage of the phase transformation. It can result in regular or chaotic pattern structure; typically, long-range interactions play an essential role there [11]. One of the examples of the regular pattern is a nano-composite permanent magnet ALNICO [12] and Sm(CoFeCuZr)$_{7.5}$ [13] where the basket-weave microstructure arises as a result of spinodal decomposition of an alloy [10]. The regular pattern structure can correspond to a stable or metastable state of a system as it takes place in the cases of stripe magnetic domains [14,15,16,17,18], "polytwinned" antiphase domain structures in tetragonal ordered alloys [19,20], lath martensite [21,22,23] or perlite in steel [24]; the latter is a particular case of the well-known eutectoid decomposition [25]. The stabilization mechanisms for the regular nanoscale microstructures are related to the tendency to minimization of the energy of long-range interactions. Morphology of the forming structure depends on external fields [26,27] and their gradients [27,28], as well as on the dissipation of the stored energy, by plastic relaxation in the case of elastic interaction [22,23]. There are even some rigorous mathematical results on the formation of stripe and checkboard patterns in Ising model with the long-range interactions included [29]; the computational simulations see, e.g., in Ref. [30].

Here we focus on the other class of materials which attracts increasing attention in last decades. In this case, the microstructure contains stable or long-living metastable nanosize-scale precipitates embedded in a host [31]. Such heterogeneous state is typical for the so called nanoscale granular materials [32] and was observed in many technologically important alloys. Examples include pre-Guinier-Preston zones [33] (or K-state [21]) and Guinier-Preston-Bagaryatsky zones in aluminum alloys [34,35] which are supposed to play a decisive role in their strengthening [36], heterophase fluctuations [37] resulting in athermal omega-phase in Ti- and Zr-based alloys [38,39,40] and in Cu-Zn system [41], precipitates of Co in Cu [42], Cu in Fe [43], Pb in Al [44].

The common feature of these structural states is their stability at a moderate temperature, the precipitates are neither grow to a macroscopic size nor disappear. The conventional theory of structural transformations in solids has difficulties explaining such states, which lies very deeply; actually, it follows just from the separation of free energy of multiphase system into bulk and surface contributions [10,21]. This assumption is by no way self-evident since a coherent precipitate of a new phase creates long-range deformations in the host. At a phenomenological level, possible violation of extensivity of thermodynamic quantities at the nanoscale was discussed in Ref. [4]. However, as was shown by Eshelby [45] the energy of coherent precipitates, at least, for the inclusion of the ellipsoidal shape, is proportional to the volume and therefore it is just a renormalization of the bulk contribution of the energy.

Coherent conjugation when crystal structures of the inclusion and the host are matched at the interface by a small homogeneous deformation is typical for small inclusions. With the inclusion growth, a character of conjugation is changed and the coherence is lost for large enough inclusions; in this case, tangential component of the deformation field is no more continuous at the interface. In this work, we demonstrate that for partially coherent precipitate a special situation takes place. To provide the partially coherent state one has to introduce topological defects, namely, misfit dislocations [21,46], and their interaction energy turns out to be different from both bulk and surface contribution to the total energy. This additional energy results in a stabilization of the nanosize precipitate.

# Two examples of nanoscale microstructures

Before considering physical mechanisms of stabilization of the nansocale microstructures, we discuss in a bit more detail two particular examples, to make the problem clearer. These examples are practically important and show most typical features of the loss of coherence by precipitates.

**A Athermal omega-phase formation**

The athermal omega-phase is observed in numerous titanium and zirconium alloys at the quenching after homogenizing annealing at temperatures corresponding to stable beta-phase (bcc) [40]. It is observed only as precipitates (and never as a bulk phase) by a displacive mechanism that transforms the structure from bcc (beta) to hexagonal (omega) via a collapse of the {111} planes of the parent bcc phase (Fig. 1). The athermal omega precipitates are typically considered to inherit the composition of the parent beta matrix. As a rule, omega-phase in titanium alloy with beta-stabilizing elements such as V, Cr, Mn, Fe, Co and Ni appear in the form of fine ellipsoidal particles with long axe along <111> and distributed uniformly over the whole of the grain volume [38].

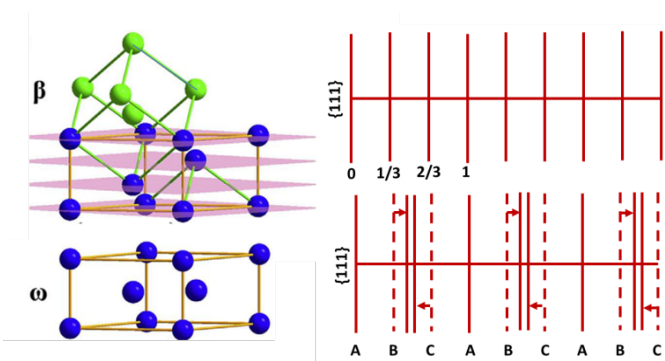

Fig. 1. Relation between bcc and omega phase crystal lattices (left) and β – ω transformation by collapse of certain {111} planes (right).

The athermal omega-phase provides us an example of a precipitate coherently conjugated with the host crystal lattice despite essential differences in the crystal structures. The truly coherent conjugation corresponds to the so called commensurate athermal omega-phase. This situation is quite exclusive (it happens in elemental Ti, Zr and Hf under pressure), the incommensurate athermal omega phase is much more typical [40], the omega-beta mismatch (that is, amplitude of displacements of the {111} planes) depends on the concentration of the dopant. Such a conjugation is semi-coherent and includes appearance of domain boundaries or topological defects as suggested in Ref. [47].

On subsequent isothermal annealing, coarsening of the omega-phase precipitates is accompanied by the diffusional partitioning of the alloying elements. As a result, the isothermal omega-phase is formed with a reduced concentration of the alloying elements and a larger lattice mismatch that finally leads to the loss of coherency.

Thus, the formation of the omega-phase precipitates results from concurrent compositional and structural instabilities of titanium- and zirconium-based alloys [48]. The detailed theory of formation of the athermal omega-phase is still absent. One can assume that the peculiar structural state of beta+omega titanium based alloys is responsible for their anomalous electronic properties such as negative temperature coefficient of resistivity [5] and provides an efficient mechanism of giant ultrasound attenuation [49] observed in these alloys [50].

**B Guinier-Preston zones and their structural features**

Though AlCu-based alloys have been discovered more than 100 years ago, still nowadays they are of great importance for light-weight constructions [36,51] such as fuselage of aircrafts (including new Airbus A380). The prominent properties of AlCu-based alloys (mainly AlCuMg) are their low specific weight combined with hardness and tensile strength which is comparable to those of steels. This hardening of Al is crucially dependent on coherent meta-stable precipitates formed at the annealing at the temperatures corresponding to the equilibrium solid solution with further quenching and aging at room or moderately high temperature. In binary AlCu alloys these precipitates are thin Cu platelets of a few nanometer thickness on the {100}-planes in fcc-Al; they are called Guinier-Preston zones (GPZ) [34].

Tempering of AlCu alloys above room temperature leads to the growth of GPZ and their transformation which includes several steps (Fig.2): GPZ I → θ'' particles (GP II zones) → θ' phase particles → θ phase particles. The GPZ I zones are just one Cu layer in {100} plane (Fig. 2d). The θ '' particles contain two or more {100} Cu layers separated by three aluminum planes. GPZ I and II are precipitates coherently conjugated with the host. The θ' particles are larger and semicoherent with the Al host, that is, they are conjugated with the host via formation of misfit dislocations [52]. Finally, the θ particles are inclusions of a thermodynamically stable phase $Al_2Cu$ incoherent with the host. The highest strength of the alloy is reached just before the precipitates loss coherency, when the compensation of long-range internal stresses takes place. Despite a long history of investigations of GPZ the key questions on the mechanisms providing their stabilization and, thus, unique mechanical properties of Al-based alloys are still open. In particular, it is unclear, what is the exact role of quenched-in vacancies. It was suggested [35] that the latter are crucially important on early stage of GPZ formation because they provide relaxation of the strain due to size mismatch between the host and solvent atoms. However, the relative concentration of quenched-in vacancies and solute atoms is just about 1/1000 [36]. It seems to be too small to affect the structure of precipitates but can accelerate essentially diffusion processes providing a decomposition of homogeneous alloy at room temperature. This is a common believe that tiny platelet shape of GPZ provides the gain in the energy on coherent strain [35]. However, to our knowledge, possible (meta)stability of this state has never been really demonstrated. Such demonstration is the main aim of this paper.

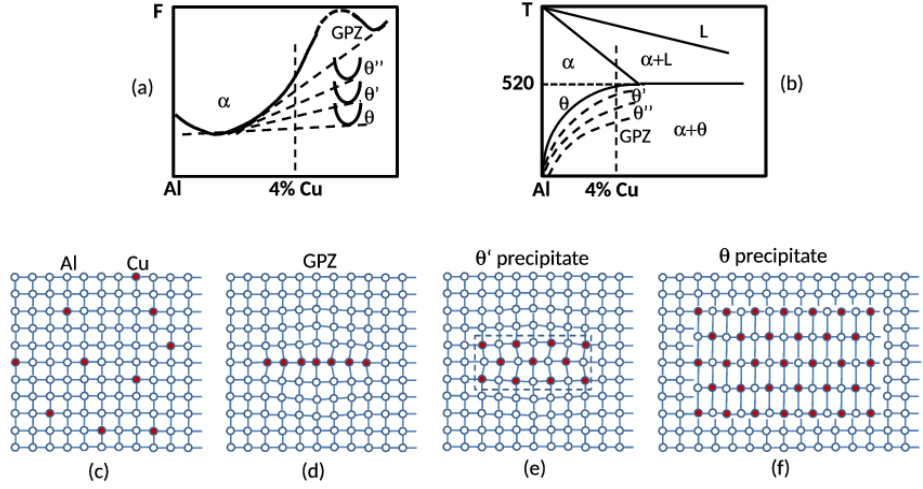

Fig. 2. Precipitation in Al-Cu supersaturated solid solution includes several steps (GPZ → θ″ particles (GP II zones) → θ′ phase particles → θ phase particles) which follow each other with increasing temperature. Schematic relation are shown free energies (a) and corresponding phase diagram (b), crystal structure of solid solution (c), GPZ (d), θ′ phase (e) and θ phase (f).

## Coherent inclusion problem

The concept of coherent conjugation is crucially important for the rest of our paper. Incoherent conjugation corresponds to coexistence of two or more macroscopically large pieces of different phases. Under such condition, each phase has a structure minimizing its chemical potential and the role of interface energy in the total energy balance is negligible. For the case of nanoscale inclusions it can be more important to minimize the interface energy than bulk energy of the inclusion. It requires an optimization of the interface to avoid energetically expensive jumps in atomic positions and to make the displacement field across the interface as smooth as possible. This kind of conjugation is called coherent.

Early stage of precipitation results in the appearance of particles coherently conjugated with the host; their structure can be, generally speaking, different from that of the corresponding bulk phase. In particular, the lattice parameters of the athermal omega-phase are essentially different from those in the equilibrium (e.g., at high pressures) thermal omega-phase [40] and the structure of GPZ is essentially different from the equilibrium $Al_2Cu$ phase [52]. The coherent long-range deformation in the host created by the particle modifies the conditions of phase equilibrium in the system, the former are now dependent on size and shape of the precipitates [53,54]. Such a picture is valid for small enough particles. When their size grows above a typical crossover size determining by the energy balance between bulk differences of chemical potentials and elastic energy of the deformations we reach the true thermodynamic phase equilibrium (as required by Gibbs conditions) which does not depend on the particles size if one neglects the surface (interface) energy. The latter case corresponds to spinodal or binodal lines of equilibrium at the phase diagram [21] while under coherent strain a new phase can be reached at larger overcooling. In this section, we consider an isolated particle coherently conjugated to the host. In the next section the system under the near-crossover conditions will be considered when a semicoherent conjugation of the precipitates with the host.

Eshelby's solution [45] for an ellipsoidal inclusion in an infinite homogeneous isotropic elastic material played a key role in understanding of the precipitation. As was shown by Eshelby the energy of elastic strains created by an ellipsoidal coherent inclusion of a new phase in the host has the form

$$E_{el} = -\frac{1}{2} V \sigma_{ij}^{inc} \varepsilon_{ij}^{t} ,\qquad(1)$$

where $V$ is the volume of the inclusion, $\varepsilon_{ij}^{t}$ is the transformation strain in unconstrained inclusion, that is, in the bulk new phase without the host, $\sigma_{ij}^{inc}$ is the elastic stress inside the inclusion which turns out to be homogeneous in the case of the ellipsoidal shape:

$$\sigma_{ij}^{inc} = C_{ijkl}(S_{klmn}\varepsilon_{mn}^{t} - \varepsilon_{kl}^{t})\qquad(2)$$

$S_{ijkl}$ is the so called Eshelby tensor [45] connecting deformations inside the inclusion in constrained (c) and unconstrained (t) conditions, $\varepsilon_{ij}^{c} = S_{ijkl}\varepsilon_{ij}^{t}$. Since the approach by Eshelby is based on the classical continual elasticity theory, the tensor $S_{ijkl}$ for the case of the ellipsoidal inclusion depends only on Poisson's ratio of the material and the aspect ratios of the main axes of the inclusion [45]. As a result, this approach does not take into account the inclusion-size effect on elastic behavior exhibited by particle-matrix composites (e.g. [55,56,57]), and coherent stresses will lead only to a shift of the equilibrium line at the phase diagram renormalizing the free-energy difference.

This limitation has motivated studies of Eshelby-type inclusion problems using extensions of the classical elasticity theories, which contain material length scale parameters [58] or the curvature of interface [59,60]. These approaches predict size-dependent elastic strain for inclusions of few nanometers and were applied to investigation of the effect of lattice mismatch on properties of nanostructures such as buried quantum dots [59,61], nanowires [62] and composites [63].

Here we follow the traditional approach by Eshelby and consider another mechanism of the size-dependent behavior suggested at the qualitative level in Ref. [54], namely, a crossover from coherent to incoherent conjugation of the inclusion with the host. As was shown in Ref. [21], minimum of the elastic energy is reached for the coherent inclusion with the shape of a narrow plate. Therefore we restrict ourselves by the consideration of the penny-like-shape inclusion, a typical case for the inclusions with a large lattice mismatch and small surface tension [21].

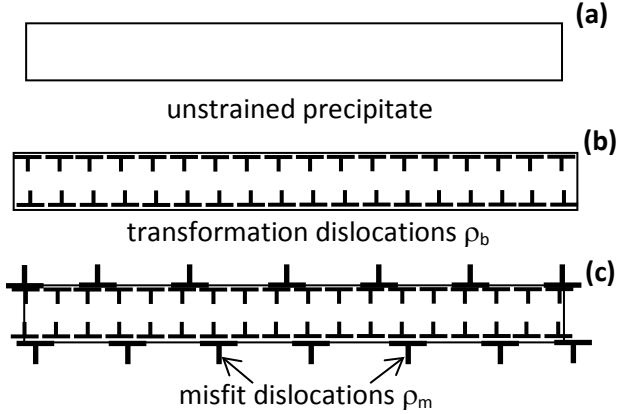

Рис. 3. A plate (penny-like-shape) inclusion (a) containing virtual dislocations (b) with linear density $\rho_b$ determining the transformation deformation and misfit dislocations (c) with the linear density $\rho_m$.

We will use the dislocation model of the conjugation of inclusion with the host; its equivalence to the Eshelby approach was discussed in Ref. [64]. Namely, we assume that the conjugation is accompanied by appearance of continuously distributed dislocations and/or dislocation loops at the interface (Fig. 3). If the total tensor of intrinsic deformations created by such defects is equal to the intrinsic deformation tensor of the inclusion this, of course, provides also equality of strains. The network of virtual dislocations is characterized by the tensor dislocation density

$$\rho_{ij}^b = \rho_\xi \xi_i b_j \tag{3}$$

where $\rho_\xi$ is the linear density of dislocations with the axes along the direction $\xi$ with the Burgers vector **b**. The quantity (3) is related to the jump of plastic distortion at the interface $\varepsilon_{ij}^t$ via Bullough-Bilby connection [65]

$$\rho_{ij}^b = -e_{ikl} n_k \varepsilon_{lj}^t \tag{4}$$

where $e_{ikl}$ is the unit antisymmetric tensor and **n** is the normal vector at the interface. We assume that the inclusion has a shape of a plate with **n** parallel to z axis (z=$x_3$) and that the deformation is tetragonal, that is, $\varepsilon_{11}^t = \varepsilon_{22}^t = \varepsilon^t$. Then, the only non-vanishing components of the tensor $\rho_{ij}^b$ are $\rho_{12}^b = \rho_{21}^b = \rho_b b = \varepsilon^t$, and the elastic field created by the inclusion can be attributed to a superposition of two families of mutually orthogonal edge dislocations; one of them is schematically shown in Fig. 3b.

To determine elastic energy of the coherent inclusion we have to take into account interaction energies and each dislocation family and between the families. Taking into account that the Burgers vectors have opposite signs at the lower and upper surfaces of the plate we obtain:

$$\begin{aligned} E_{el}^c &= 4L \frac{\mu b^2}{2\pi(1-\nu)} \rho_b^2 \int_{-L}^{L} \ln\left(\frac{\sqrt{(x-x')^2}}{\sqrt{(x-x')^2 + d^2}}\right) dx dx' \\ &= V \frac{\mu(\varepsilon^t)^2}{(1-\nu)} \left(1 + \frac{d}{2\pi a}\left(2\ln\left(\frac{a}{d}\right) + 3\right)\right) \end{aligned} \tag{5}$$

where $d$ is the thickness of the plate, $a = 2L$ is its in-plane size, $V = da^2$. The integrand in Eq.(5) is nothing but the interaction energy of parallel dislocations per unit length. This interaction is long-ranged, with logarithmic dependence of potential on the distance. The final expression in the right-hand side of Eq.(5) agrees with that derived in Ref. [21] by another way. It allows a simple interpretation, namely, the first term (with 1) is the energy of a homogeneous elastic field inside the plate and the second one (proportional to $d/a$) is the energy of the loop of an effective dislocation along the perimeter of the inclusion. For this coherent inclusions $d \ll a$ and the last term can be neglected, in agreement with the conclusion of Eshelby that the elastic energy is proportional to the volume of the inclusion. Note that the expression (5) is valid, except numerical factors of the order of one, for any shape of the inclusion [21] assuming that its transverse size $L$ is much larger than its thickness $d$.

## Coherent to incoherent crossover

The simplest model describing a transition from coherent to incoherent regime is the Frenkel-Kontorova model of an atomic chain lying at incommensurate substrate [46,66,67]. In this model, incoherent regime corresponds to the case when elastic energy is much large then the energy of interaction with the substrate so that the atomic chain remains undeformed. Oppositely, in the case of very strong interaction energy the atomic chain takes the interatomic distance of the substrate (coherent regime). When both contributions are relevant the semicoherent regime arises. In this case the chain consists of coherent pieces interrupted by topological defects (solitons). One can assume, the in three-dimensional case the transition between coherent and incoherent regimes also involves a superlattice of topological defects, the key idea for further consideration.

The mechanism of partial coherent strain compensation by an array of misfit dislocations was suggested long ago by Nabarro [68] (see also recent review [69]). In that approach, the effect of misfit dislocations was reduced to a local renormalization of lattice mismatch. Below we develop this idea by a straightforward calculation of different contribution to the total energy. Contrary to Refs. [68,69] we do not take into account possible locking of the dislocations by Peierls relief since in close packed metallic systems such as Al-Cu the latter in known to be negligible [70].

We assume that the semicoherent conjugation between inclusion and host is carried out by a formation of additional network of misfit dislocations at the interface (Fig. 3c) with the Burgers vectors opposite to those of virtual dislocations (4) which are responsible for the transformation deformation. In terms of Ref.[23] these dislocations provide the plastic part of misfit strain relaxation. An additional contribution to the elastic energy arises, due to interaction within families of misfit dislocations:

$$E_{el}^m = -4L \frac{\mu b^2 2N}{2\pi(1-\nu)} \left( \ln\left(\frac{r_0}{d}\right) + \sum_{k=1}^{N} \left( \ln\left(\frac{1}{d^2+p^2k^2}\right) - \ln\left(\frac{1}{p^2k^2}\right) \right) \right) \quad (6)$$

where we have separated explicitly the term $\propto \ln(r_0)$ determining self-energy of dislocations and renormalizing the surface tension, $p = 1/\rho_m$ is the distance between misfit dislocations, $\rho_m$ is their linear density and $r_0$ is a cutoff parameter (dislocation core radius) [70]. Calculating the sums in Eq.(6) and passing to the limit $N \to \infty$ we have:

$$E_{el}^m = \frac{\mu b^2 4L^2}{\pi(1-\nu)p} \left( \frac{\pi d}{p} + \ln\left(1 - \exp\left(-\frac{2d\pi}{p}\right)\right) - \ln\left(\frac{2\pi d}{p}\right) - \ln\left(\frac{r_0}{d}\right) \right) \quad (7)$$

Apart from the contribution $E_{el}^m$ one has to take into account the interaction energy between misfit dislocations and virtual dislocations responsible for the transformation:

$$E_{el}^{mb} = 8L \frac{\mu b^2}{2\pi(1-\nu)} \rho_b b N \int_{-L}^{L} \ln\left(\frac{\sqrt{x^2}}{\sqrt{x^2+d^2}}\right) dx$$

$$= -\frac{\mu b^2 V}{(1-\nu)} 2\rho_b \rho_m \quad (8)$$

The total elastic energy is the sum of the contributions (5), (7) and (8): $E_{el}^{sc} = E_{el}^c + E_{el}^m + E_{el}^{mb}$. Choosing $b$ as the unit length, $\mu/(1-\nu)$ as the energy density unit and taking into account that $\rho_b = \varepsilon^t/b$, $\rho_m = \varepsilon^m/b$ ($\varepsilon^m$ is the average deformation created by misfit dislocations, its value is determined by the minimization of $E_{el}^{sc}$), the elastic energy per unit area can be represented as

$$\tilde{E}_{el}^{sc} = \frac{E_{el}^{sc}(1-\nu)}{\mu b a^2} = d(\varepsilon^t - \varepsilon^m)^2 + \frac{(\varepsilon^t)^2 d^2}{2\pi a}\left(2\ln\left(\frac{a}{d}\right)+3\right)$$
$$+ \frac{\varepsilon^m}{\pi}\left(\ln\left(1-\exp(-2\pi d\varepsilon^m)\right) - \ln(2\pi d\varepsilon^m) + \ln(d) + e_0\right) \quad (9)$$

where the dimensionless parameter $e_0 = \ln(r_0/b)$ depends on details of the structure of the core of misfit dislocations [71] and lies within the limits $0 < e_0 < 1$. It is proportional to the core energy of the dislocation.

The expression (9) has a transparent physical meaning. The first contribution in the right-hand side coincides with the coherent strain energy (5), with the lattice mismatch decreased by the value $\varepsilon^m$ by introducing misfit dislocations. The second contribution proportional to $d/a$, as in Eq.(5), gives the energy of deformations created by the inclusion edges. It becomes essential if $\varepsilon^m \approx \varepsilon^t$ which suppresses the bulk contribution. At last, the third term proportional to the density of misfit dislocations describes the renormalization of the surface tension. The density of the misfit dislocations $\rho^m$ and the corresponding deformation $\varepsilon^m$ minimizing the energy $\tilde{E}_{el}^{sc}$ for the case of thin plate ($d \ll a$) depend only on $d$ and $e_0$.

Fig. 4 displays the dependence of the ratio $\varepsilon^m/\varepsilon^t$ on $d$ obtained by numerical minimization of Eq.(9) for different $e_0$. For large misfit $\varepsilon^t$ = 0.1 and small core energy (small $e_0$) formation of the misfit dislocations and, thus, transition from coherent to semicoherent inclusion is energetically favourable for any $d$. In this case, the ratio $\varepsilon^m/\varepsilon^t$ remains close to 1 (the curves 1,2 in Fig. 4a) reaching minimum at $d \approx b$, and misfit dislocations compensate lattice mismatch between precipitate and host phases almost completely. At the increase of the parameter $e_0 > 0.5$ or decrease the value of misfit $\varepsilon^t$ the behaviour $\varepsilon^m(d)$ is changed qualitatively and the formation of the misfit dislocations is possible only for $d$ larger than some critical value $d_{cr}$ (the curve 3 in Fig. 4a and Fig. 4b).

The curves in Fig.4a correspond to the case of quite large misfit; its decrease results in the growth of $d_{cr}$ (Fig.4b). The issue on the critical inclusion size corresponding to the loss of coherence has been discussed already [72,73,74] by analysing the energetics of creation of the first dislocation loop. The value of $d_{cr}$ as calculated here is smaller (approximately twice) than predicted by Brooks criterion $d_{cr} \approx b/2\varepsilon^t$ [72] and depends on the core energy $e_0$. For large misfit and small enough values of $e_0$ the model predicts a qualitatively different behaviour in comparison with the previous considerations. In this case, a gradual loss of coherence takes place with the inclusion growth. It is a "soft" scenario in contrast with "hard" coherent - semi-coherent crossover which realizes for small misfit and/or larger core energy for $d > d_{cr}$.

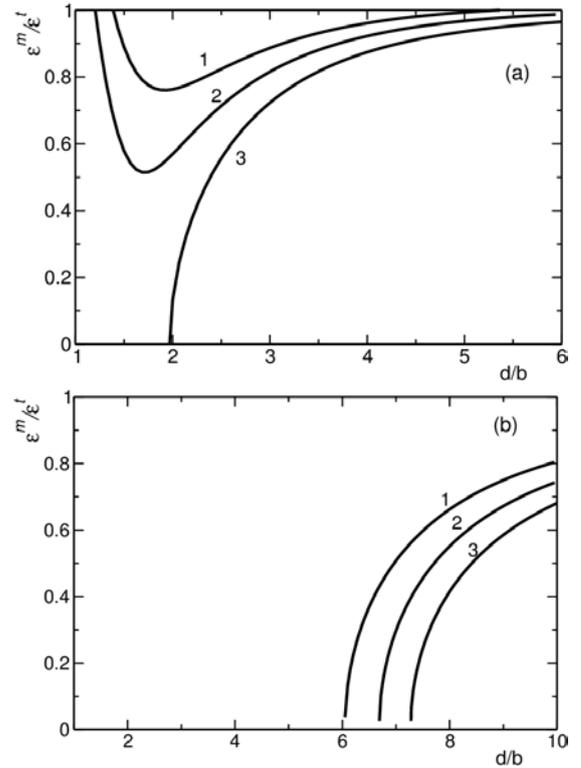

Рис. 4. Relative density of misfit dislocations $\varepsilon^m/\varepsilon^t$ as a function of the thickness $d$ for different misfit and dislocation core parameters. (a) $\varepsilon^t$ =0.1, $e_0$ = 0.45, 0.5, 0.55 for the curves 1, 2, 3, respectively. (b) $\varepsilon^t$ =0.05, $e_0$ = 0.1, 0.2, 0.3 for the curves 1, 2, 3, respectively.

## Stabilization of nano-inclusion due to misfit strain

Total formation energy of the plate inclusion has the form

$$E_{inc} = da^2 \Delta f + 2\sigma a^2 + E_{el}^{sc}, \quad (10)$$

where the first term is a chemical contribution, the second one is the surface energy and the last one is the elastic energy discussed above. The difference of the free energies per unit volume for the competing phases, $\Delta f$, is determined by the change in the chemical bonding due to variation of composition or/and the crystal structure, as well as by the change of entropy. Near the temperature of phase equilibrium $T_c$, the value $\Delta f$ is usually proportional to the overcooling temperature $\Delta T = T - T_c$ [75]. Substituting Eq.(9) into Eq.(10) one can represent the inclusion energy per unit area as

$$\widetilde{E}_{inc} = d\left[\left(\varepsilon^t - \varepsilon^m\right)^2 - \Delta \widetilde{f}\right] + \frac{\left(\varepsilon^t\right)^2 d^2}{2\pi a}\left(2\ln\left(\frac{a}{d}\right) + 3\right)$$
$$+ \frac{\varepsilon^m}{\pi}\left(\ln\left(1 - \exp\left(-2\pi d \varepsilon^m\right)\right) - \ln\left(2\pi d \varepsilon^m\right) + \ln(d) + e_0\right) + 2\widetilde{\sigma} \quad (11)$$

where $\widetilde{E}_{inc}, \Delta \widetilde{f}$ and $\widetilde{\sigma}$ are the energy densities multiplied by $(1-\nu)/\mu b a^2$, $(1-\nu)/\mu$ and $(1-\nu)/\mu b$, respectively. Fig. 5 shows the inclusion energy as a function of $d$ for the cases of "hard" (a) and "soft" (b) coherent - semi-coherent crossover. The curves 2,2' corresponds to the incoherent inclusion when elastic strains are absent. Its slope is determined by the change of the free energy $\Delta f$ at the formation of the new phase and vanishes at the temperature of the phase equilibrium. The curve 1 presents the elastic energy $\widetilde{E}_{el}^{sc}(d)$. In the hard case (Fig. 5a) the function $\widetilde{E}_{el}^{sc}(d)$ changes linearly while $d<d_{cr}$, with the slope determining by the energy of coherent strain. In the soft case the loss of coherence happens gradually and the slope $\widetilde{E}_{el}^{sc}(d)$ curve (Fig. 5b) decreases. An overcooling is necessary in the both case to initiate decomposition in the presence of coherent strain.

The total energy of inclusion results from competition of elastic and chemical contributions (including surface energy). For the hard crossover (Fig. 5a) the function $\widetilde{E}_{inc}(d)$ varies linearly at $d<d_{cr}$ and change slope after the loss of coherence (curve 3). In the case of small overcooling (small slope of curve 2) the energy $\widetilde{E}_{inc}(d)$ increase in dependence on $d \approx d_{cr}$ (curve 3') and inclusion is unable to grow normal direction. On the other hand, for the soft crossover (Fig. 5b) the total energy depends non-monotonously on $d$, regardless of overcooling. Thus, to transform the thin plate to the thick one it is necessary to overcome the energy barrier for the cases of soft crossover. In these situations as well as for hard crossover and small overcooling one can expect a formation of metastable plate precipitate.

Let us discuss now a relation of our model to GPZ in Al-Cu alloys. GPZ-I exists at the temperatures T < 200$^0$C are just monolayers of Cu characterized by large lattice mismatch $\varepsilon^t$ =0.1; at T > 200$^0$C GPZ-I are solved and the θ'-phase is formed instead [76]. According to the results of ab-initio calculations [77] the energy gain at the formation of GPZs is about 0.01-0.02eV/at, which agrees well with temperature range of their existence and corresponds to $\Delta \widetilde{f} = 0.004 \div 0.008$. A reasonable estimate of the energy of the coherent interface is about 0.1-0.2 J/m$^2$ which corresponds to $\widetilde{\sigma} = 0.001$. According to our consideration this is within the soft regime when increase of thickness of GPZ-I is energetically unfavourable due to generation of misfit dislocations. On the other hand, θ'-phase arising at higher temperatures and corresponding to the chemical composition Al$_2$Cu has smaller lattice mismatch $\varepsilon^t$ =0.02 which corresponds to our hard regime. Contrary to the case GPZ-I, in this case thin plates are not stable (curve 3 in Fig. 5a). Indeed, the thickness θ'-phase plates increases rapidly during a continual heating [76].

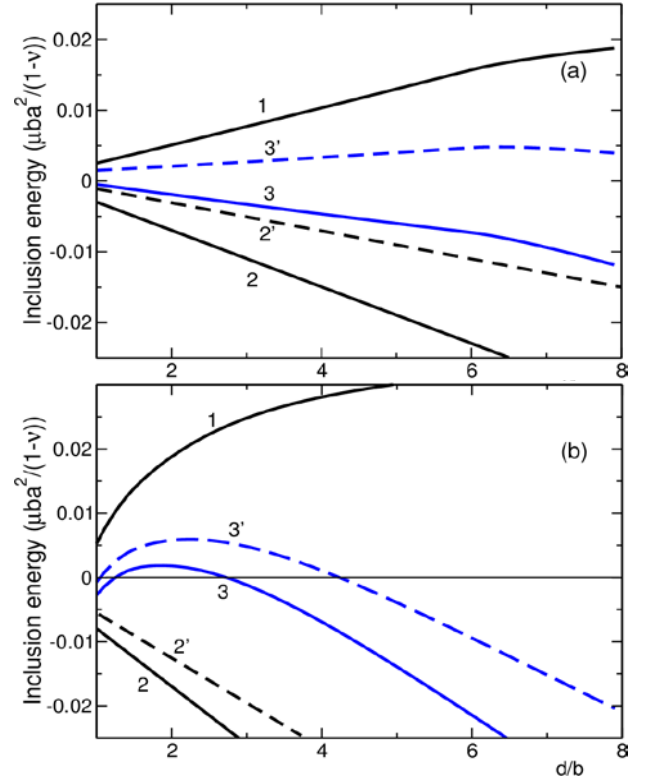

Fig. 5. The energy of plate inclusion $\widetilde{E}_{el}^{sc}$ for the cases of "hard" (a) and "soft" (b) coherent to semicoherent crossover. Curve 1 is the elastic contribution, curve 2, 2' is the chemical contribution and curve 3, 3' is the total energy. The parameters are: $\varepsilon^t = 0.05$, $\widetilde{\sigma} = 0.001$, $e_0$ = 0.1, $\Delta \widetilde{f} = 0.004$ and 0.002 for curves 2,3 and 2',3' (a), $\varepsilon^t = 0.1$, $\widetilde{\sigma} = 0.001$, $e_0$ = 0.45, $\Delta \widetilde{f} = 0.009$ and 0.007 for curves 2,3 and 2',3' (b).

Local energy minimum corresponds to $d$ = 0 which means, within our model based on continual elasticity theory, a few interatomic distances. This means also stability of completely coherent inclusion with zero density of misfit dislocations. Our model always predicts either atomically thin coherent plates or unrestricted incoherent growth of the new phase. Experimentally, in all the cases GPZ do have thickness of the order of interatomic distances and are completely coherent.

Of course, our model is oversimplified (using continual elasticity, the free energy difference and surface tension are supposed to be independent on the size of precipitate) and neglects some details which are important for the quantitate description. The main aim is to demonstrate mechanisms stabilizing atomically thin plate rather than to provide a complete quantitative theory of metastable precipitates in real Al-Cu alloys. In the latter, the structural evolution is accompanied by changes in chemical composition and crystal lattices of the inclusion phases (see Fig.2). However, it should be noted that dislocation arrays is not the only way for semicoherent conjugation of inclusion and host and other topological defects may be involved. In particular, as was shown in Ref.[78], conjugation of precipitates of topological close packed (Frank-Kasper) phases may be provided by a network of structural disclinations rather than dislocations. Formation of such precipitates in W(Mo)-Re alloys can be an explanation of anomalous solubility of interstitial impurity and improvement of mechanical properties (rhenium effect) [79,80].

## Conclusions

To summarize, energetics of partially coherent precipitate turns out to be quite peculiar. In both limiting cases, namely, incoherent precipitate and completely coherent precipitate [9] the total free energy can be represented as a sum of bulk and surface contributions which are proportional to the volume of the precipitate and to the area of interface between the precipitate and the host, respectively. This is not surprising since in both these limits there is no need to introduce any topological defects which can make the story much more complicated [78]. These defects are unavoidable in the partially coherent case. Due to a long-range character of interaction between the topological defects, their interaction energy may have unusual dependence on geometry of the system which, as we have shown, may result in the stabilization of the precipitate and determine its equilibrium shape. For the future, it would be important to generalize our quite simple model to the case of precipitates of arbitrary shape. In particular, contrary to the case of Al-Cu, for Al-Zn-Mg alloys spherical shapes of the inclusions are typical [81]. As was mentioned above, inclusions of athermal omega-phase in Ti and Zr based alloys are usually ellipsoidal [38,39]. Note also that in this paper we discuss only stabilization of the isolated precipitate; these "topologically generated" interactions can be important also for interaction between the precipitates and thus to the formation of mesostructure of inhomogeneous alloys.

## Acknowledgements

MIK acknowledges a financial support by ERC Advanced Grant No. 338957 FEMTO/NANO and by NWO via Spinoza Prize. YNG acknowledges financial support from the Russian Science Foundation (grant 14-12-00673).